\documentclass[english,twocolumn,aps,prb,longbibliography,superscriptaddress]{revtex4-2}
\usepackage{babel}
\usepackage{amsmath,mathtools}
\usepackage{amssymb}
\usepackage{graphicx}
\usepackage{tabularx}
\usepackage{xcolor}
\usepackage{hyperref}
\hypersetup{
	colorlinks,
	citecolor=green,
	linkcolor=red,
	urlcolor=magenta}
\begin{document}
\title{Valley Hall Effect and Kink States in Topolectrical Circuits }
\author{S M Rafi-Ul-Islam }
\email{e0021595@u.nus.edu}
\selectlanguage{english}%
\affiliation{Department of Electrical and Computer Engineering, National University of Singapore, Singapore 117583, Republic of Singapore}
\author{Zhuo Bin Siu}
\email{elesiuz@nus.edu.sg}
\selectlanguage{english}%
\affiliation{Department of Electrical and Computer Engineering, National University of Singapore, Singapore 117583, Republic of Singapore}
\author{Haydar Sahin}
\email{sahinhaydar@u.nus.edu}
\selectlanguage{english}%
\affiliation{Department of Electrical and Computer Engineering, National University of Singapore, Singapore 117583, Republic of Singapore}
\affiliation{Institute of High Performance Computing, A*STAR, Singapore 138632, Republic of Singapore}
\author{Mansoor B.A. Jalil}
\email{elembaj@nus.edu.sg}
\selectlanguage{english}%
\affiliation{Department of Electrical and Computer Engineering, National University of Singapore, Singapore 117583, Republic of Singapore}
\begin{abstract}
We investigate the emergence of topological valley Hall and kink states in a two-dimensional topolectrical (TE) model as a result of broken chiral and reflection symmetries. The TE system consists of two segments hosting distinct topological states with opposite signs of the valley Hall index, and separated by a heterojunction. In the practical circuit, the valley Hall index can be flipped between the two segments by modulating the onsite potential on the sublattice nodes of the respective segments. The presence of resistive coupling, which introduces non-Hermiticity in the system, subsequently leads to the emergence of gapped and gapless valley and kink states in the admittance spectra. These topological modes can be detected electrically by the impedance readouts of the system which can be correlated to its admittance spectra. Finally, we confirm the robustness of the valley Hall and kink states via realistic LTspice simulation taking into account the tolerance windows and parasitic effects inherent in circuit components. Our study demonstrates the applicability of TE circuit networks as a platform to realize and tune valley-dependent and kink topological phenomena.
\end{abstract}
\maketitle
\section{Introduction} 
The study of topological phases has emerged as a frontrunner topic in condensed matter physics owing to the unconventional properties of such phases \cite{roy2009topological,gomes2012designer,pollmann2012symmetry}, which possess a non-trivial band topology. The valley is meanwhile a new degree of freedom that can be found in lattice models with various symmetries  \cite{suzuki2014valley,yao2008valley,wu2013electrical}.  This additional valley freedom is useful in various technologically significant applications such as robust electronic transport \cite{mak2014valley,ziegler2006robust}, energy propagation \cite{chen2017valley}, and information processing \cite{qi2015giant,ferdous2018valley}. As a result, the valley degree of freedom has given rise to a completely new branch of technology named ``valleytronics"  \cite{schaibley2016valleytronics,vitale2018valleytronics,liu2019valleytronics} that has revolutionized many existing fields such as photonics \cite{lu2014topological}, metamaterials  \cite{krishnamoorthy2012topological}, condensed matter \cite{wen2017colloquium}, acoustics  \cite{ma2019topological}, and next-generation quantum computing \cite{kitagawa2010exploring,lu2018valley,goldman2016topological}. Breaking the inversion symmetry in a lattice model results in a valley-dependent Hall conductivity and a quantum valley Hall effect \cite{pal2017edge,pan2014valley}. The inversion symmetry can be broken by inducing alternating mass terms in the lattice Hamiltonian \cite{chen2019valley}. A domain wall-type interface is created when two lattice segments with opposite valley responses are joined together. Robust valley kink states appear at the interfaces of such heterojunctions  \cite{gao2018topologically,jung2011valley}. These novel valley kink states are useful in many promising phenomena such as Klein tunneling \cite{rafi2020strain,stander2009evidence,rafi2020topoelectrical}, anti-Klein tunneling \cite{rafi2020anti}, spin-valley locking \cite{saito2016superconductivity}, and quantum  memory \cite{lvovsky2009optical}. 

More recently, topological boundary states have been realized in many Hermitian  \cite{obana2019topological,hafezi2013imaging} and non-Hermitian systems  \cite{rafi2022critical,gong2018topological,rafi2022unconventional,esaki2011edge,rafi2022interfacial}. These topologically nontrivial boundary states are characterized by the topological index (i.e., Chern number \cite{aidelsburger2015measuring} and Hall conductivity\cite{zheng2002hall}) of their gapped bulk energy bands, and exhibit gapped and gapless states on their boundaries depending on the model parameters. These topological boundary states are protected by symmetries such as  time reversal \cite {he2016photonic} or spatial inversion symmetry \cite{casteels2017quantum},  and are robust against local perturbations and disorders. These novel topological boundary phases and valley Hall states have recently been demonstrated in different platforms such as photonics systems \cite{lu2014topological,lu2018valley,dong2017valley}, metamaterials \cite{goldman2016topological,dong2021tunable,zhou2020voltage,li2021topological}, and quantum wells \cite{bernevig2006quantum}. Although boundary states and valley kink modes promise to bring dramatic changes to existing technologies, it is difficult to realize and observe multiple topological valley and boundary phases in the same lattice model because of difficulties in  the dynamical modulation of the system parameters (e.g., fixed lattice constants and weak spin-orbit coupling). Additionally, all these platforms involve experimentally complex sample preparation, which is very vulnerable to perturbations and impurities.

Lattice arrays comprising electrical components such as inductors and capacitors known as topolectrical (TE) circuits  \cite{rafi2020realization,lee2018topolectrical,rafi2021topological,hofmann2020reciprocal,rafi2020anti,zhang2020topolectrical,rafi2022system} have become the frontier experimental testbed in the quest for alternative platforms to study different topological states. Compared to other platforms, TE circuits offer better tunability of system parameters such as the interaction strength and phases. Recently, many exotic and novel features such as edge states   \cite{olekhno2020topological,lee2018topolectrical}, corner states \cite{imhof2018topolectrical}, quantum spin Hall states \cite{zhu2019quantum,sun2020spin}, chiral magnetic effects \cite{tan2018emulating,lin2014ac}, topological photonic states \cite{xie2018second}, and nodal ring states \cite{luo2018topological,rafi2021non,luo2018nodal,li2019emergence} have been proposed in electrical circuit networks \cite{rafi2020topoelectrical,lee2018topolectrical,hofmann2020reciprocal,zhang2020topolectrical}.  The topological states depend on the connectivity between the electrical components rather than their relative locations in real space. The TE circuit models also provide better flexibility in varying the system parameters coupled with the convenient and accurate readout of system characteristics. 

In this paper, we design and propose a general framework to realize various topological valley phases and kink states based on the electrical responses in a two-dimensional TE circuit model. We have explained in detail the fundamental relationship between the admittance band structure and the impedance profiles through the circuit Green’s function in our previous works \cite{rafi2020realization,rafi2021topological,rafi2020topoelectrical}. By tuning the onsite interaction strength on the different sublattice nodes, we obtain a transition from gapless to gapped edge states in the admittance spectra.  The valley-dependent Hall conductivity is calculated using the Kubo formula and verified through the impedance spectrum. We study the valley kink states that result from cascading two TE segments with opposite signs of the valley Hall responses together. The opposite signs of the valley Hall responses can be realized by reversing the sign of the onsite capacitance, and hence the Laplacian mass term, on both sides of  heterojunction. Both gapless and gapped kink states can be obtained by varying the relative strengths of the onsite capacitance to the resistive coupling strength. The topological kink states are localized at the interface of the heterojunction and can be distinguished by their terminal impedances. Since such TE circuit models can be implemented in typical breadboards or printed circuit boards with basic electrical components, our model not only opens new experimental possibilities and directions for the realization of various topological valley phases, but also helps in the design of multifunctional valleytronic devices. 
\section{Topoelectrical Valley Circuit Model}
\begin{figure*}[htp!]
  \centering
    \includegraphics[width=0.9\textwidth]{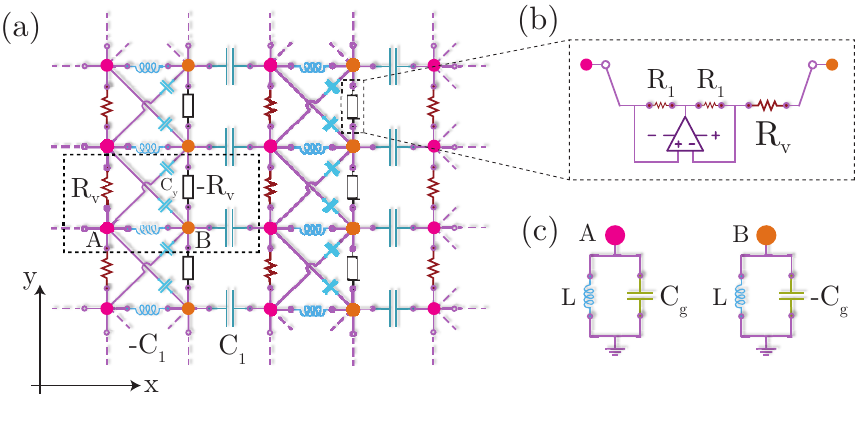}
  \caption{(a) Schematic of the valley TE lattice model that hosts valley-dependent topological phases. Here, the red and orange circles represent the $A$ and $B$ sublattice nodes, respectively. The unit cell is delineated by the dotted box.  The two alternating sublattices are connected by a capacitor $C_1$ and inductor of  $-C_1$ (i.e., an inductance of $1/(\omega^2 C_1)$) along the $x$ direction for the intracell and intercell connections respectively .   Along the $y$ direction, neighboring nodes on opposite sublattices are connected by a capacitor $C_y$ and nodes on the same sublattice connected through alternative combinations of resistive elements $\pm R_v$. (Note that there is a $\pi$ phase difference in the resistive coupling between the $A-A$ and $B-B$ connections along the $y$ direction within the same unit cell.) (b) We can make use of the negative resistance converter to realize the $\pi$ phase difference between the $A-A$ and $B-B$ connections along the $y$ direction. The combination of resistors $R_1$ and $R_v$ along with an ideal operational amplifier with supply voltages $V -$ and $V +$ acts as a negative resistance converter with current inversion. (c) Grounding mechanism of our valley TE circuit. All nodes are connected to ground by a common capacitor ($C$) and inductor ($L$). Furthermore, each $A$ and $B$ node is coupled to the ground with the same magnitude but opposite signs of the coupling strength by an onsite capacitor and an inductor, respectively.  }
  \label{fig1}
\end{figure*}

Consider the TE circuit model consisting of basic electrical components such as inductors and capacitors in Fig. \ref{fig1}. An AC current of angular frequency $\omega$ flows through the circuit, which comprises two different types of sublattice nodes labeled as the $A$ and $B$ type nodes respectively indicated as the red and orange circles in the figure. Along the $x$ direction, the two adjacent sublattice nodes within the same unit cell are connected by an inductor with an admittance equivalent to that of a capacitance of $-C_1$ (i.e., an inductance of $1/(\omega^2 C_1)$ ).   The adjacent nodes in neighboring unit cells are connected by a capacitance $C_1$ (see Fig. \ref{fig1}a). Along the $y$ direction, adjacent nodes on different sublattices are connected by a capacitance $C_y$ while nodes on the same sublattice are  connected through alternative combinations of positive and negative resistive elements $R_v =\frac{1}{i\omega r}$, where $r$ is a resistance. Note that the $A - A$ and $B - B$ couplings along the $y$ direction have a relative $\pi$ phase difference in their resistive couplings within the same unit cell. Such $\pi$ phases in the resistive elements can be obtained using negative resistance converters (NRCs) (see Fig. \ref{fig1}b), which break the reflection symmetry along the $y$ direction. Each $A$ and $B$ node is connected to the ground via a common capacitor $C$ and inductor $L$. Each $A$ and $B$ node is further grounded by another onsite potential capacitor $C_g$ and inductor $-C_g$, respectively. The common grounding inductor ($L$) is used to adjust the offset of the admittance dispersion to a common value for all nodes \cite{rafi2022critical,rafi2022interfacial,rafi2019transport}.

\section{Laplacian Phases}
The TE circuit can be described by its Laplacian, which is analogous to the tight binding Hamiltonian in quantum physics  \cite{rafi2020realization,lee2018topolectrical,rafi2020topoelectrical,zhang2022anomalous}. The Laplacian at the resonant frequency of $\omega_r=1/\sqrt{2C_y L}$ multiplied by $i\omega_r$, which we shall refer to as the normalized Laplacian for short subsequently, is given by
\begin{equation}
	\begin{aligned}	
	L_{TE}(k_x, k_y) =&  \left( -C_1+C_1\cos(k_x) +2 C_y \cos(k_y) \right) \sigma_x \\
				& + C_1\sin(k_x)\sigma_y + \left(C_g + 2 R_v \sin(k_y) \right) \sigma_z , 
	\end{aligned}
\label{WSMham}  
\end{equation}
where  $\sigma= (\sigma_x, \sigma_y, \sigma_z)$ are the Pauli matrices denoting the A/B sublattice degree of freedom.
 
The circuit hosts both topologically trivial and non-trivial phases depending on the relative magnitudes of the circuit parameters $C_g$, $R_v$, $C_1$ and $C_y$. We first investigate the transition points between the topologically trivial and non-trivial phases in the $(C_g, R_v, C_1, C_y)$ parameter space at which the eigenvalue spectrum of the Laplacian in Eq.~\eqref{WSMham} becomes gapless. This happens when the coefficients of all the Pauli matrices in Eq. \eqref{WSMham} are simultaneously zero for some real $\vec{k}$ in the Brillouin zone. For the coefficient of $\sigma_z$ to be zero, we require $k_y=\pm\sin^{-1} (C_g / R_v)$ which has a real solution when  $|R_v| \geq |C_g|$, while the  coefficient of $\sigma_y$ is zero when $k_x = 0,\pi$. Substituting $k_x=\pi$ and $k_y=\cos^{-1} (C_g/R_V)$ into the coefficeint of $\sigma_x$ in Eq.~\eqref{WSMham}, we have, $C_1 (1-\cos(k_x)) + 2C_y \cos(k_y) = 2 ( -C_1 \pm C_y\sqrt{1-(C_g/R_v)^2})$. This is zero when $(C_1/C_y)^2 + (C_g/R_v)^2 = 1$. Thus, the phase transition points occur at $(C_1/C_y)^2 + (C_g/R_v)^2 = 1$, and at $|C_g|=|R_v|$. 

\begin{figure*}[htp!]
  \centering
\includegraphics[width=0.9\textwidth]{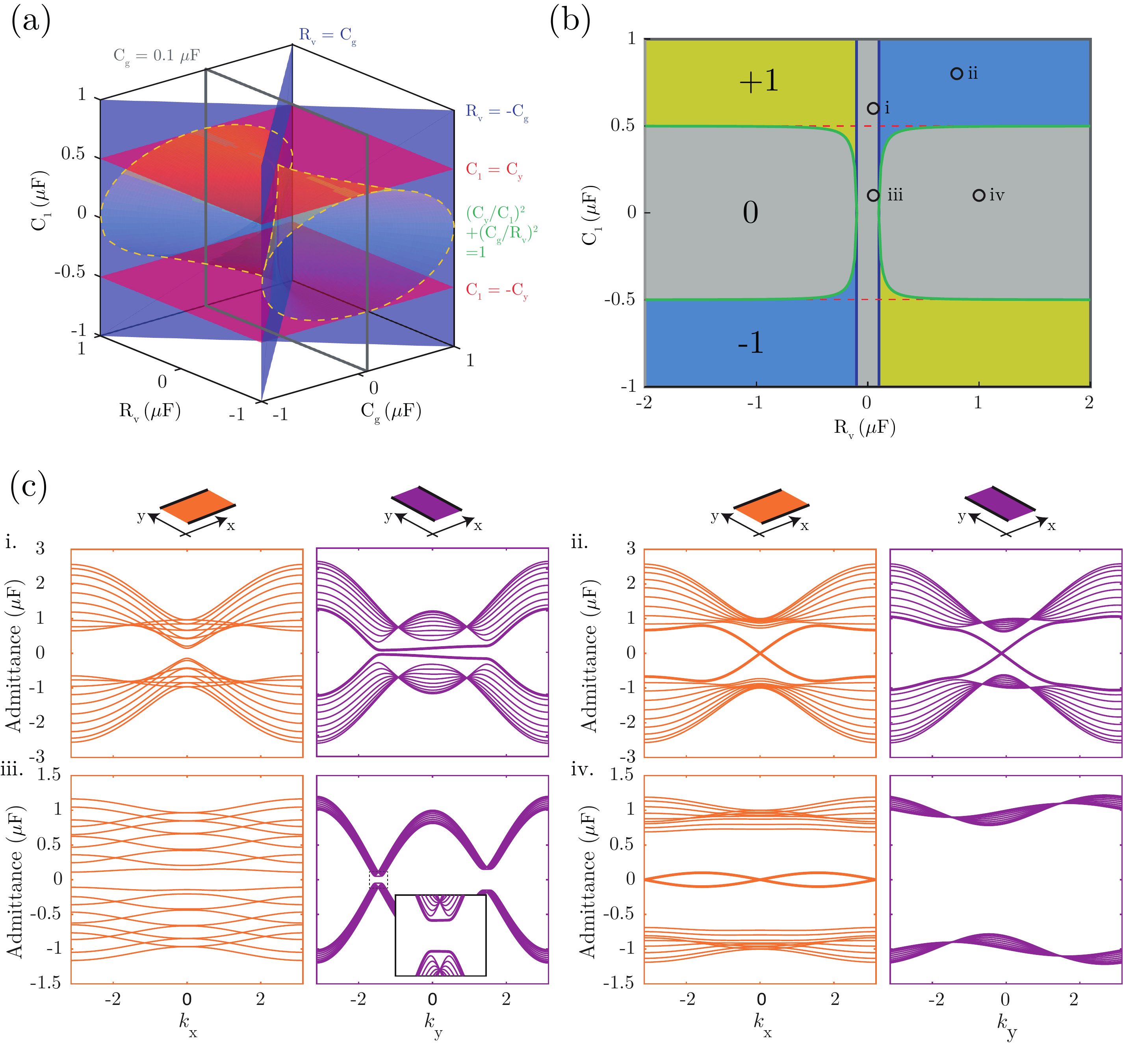}
  \caption{ (a) The phase transition surfaces in $(C_g,R_v,C_1)$ space at a fixed $C_y=0.5\ \mu \mathrm{F}$. The green surfaces outlined by the dotted lines denote the $(C_1/C_y)^2+(C_g/R_v)^2=1$ surfaces, and the blue planes the $|R_v|=|C_g|$ planes. The unfilled orange box denotes the $C_g=0.1\ \mu \mathrm{F}$ plane shown in panel (b). (b) The Chern numbers as functions of $C_1$ and $R_v$ at $C_g = 0.1\ \mu \mathrm{F}$ and $C_y=0.5\ \mu \mathrm{F}$. The gray areas denote the topologically trivial phase where the Chern number is zero. The dark blue lines denote the $|R_v|=|C_g|$ lines, and the green lines are the projections of $(C_1/C_y)^2+(C_g/R_v)^2=1$  onto the $C_g = 0.1 \ \mu \mathrm{F}$ plane. The points (i) to (iv) denote the values of $R_v$ and $C_g$ at which the dispersion relations in the finite $x$-width and finite $y$-width nanoribbon geometries are plotted in panel (c). The $(C_g,C_y,C_1,R_v)$ values of these points are (i) (0.1, 0.5, 0.6, 0.025) $\mu \mathrm{F}$, (ii) (0.1, 0.5, 0.8, 0.3) $\mu \mathrm{F}$, (iii) (0.1, 0.5, 0.1, 0.025) $\mu \mathrm{F}$, and (iv) (0.1, 0.5, 0.1, 0.5) $\mu \mathrm{F}$. (c) Admittance band dispersions of the TE model at the four $(C_1, R_v)$ points denoted in panel (b) in the nanoribbon geometry with (left) infinite length along the $x$ direction and 10 unit cells along the $y$ direction, and (right) infinite length along the $y$ direction and 10 unit cells along the $x$ direction. The nanoribbon geometries are schematically illustrated by the schematics at the top of the figure where the thick black borders at the edges denote open boundary conditions at the edges while the borderless edges extend to infinity.  The thick lines in the admittance plots denote the edge states. Note that those edge states associated with (i), (ii) and (iv) are trivial because they do not cross the band-gap, while that of (iii) are non-trivial and band-gap crossing.   }
  \label{gphasesFig}
\end{figure*}

Fig. \ref{gphasesFig}a shows the surfaces on which these phase transition points lie in the $(C_g,R_v,C_1)$ space at $C_y=1.5\ \mu \mathrm{F}$. To determine which side of the phase transition surfaces correspond to the topologically non-trivial phases, we numerically calculated the Chern numbers of the normalized Laplacian using the Fukui algorithm \cite{fukui2005chern} that  provides a numerical means of evaluating the Chern number on a discretized lattice. Fig. \ref{gphasesFig}b shows the Chern numbers at $C_y=1.5\ \mu \mathrm{F}$ and $C_g=0.1\ \mu\mathrm{F}$. The topologically non-trivial phases with finite Chern numbers occur at the intersections of $(C_1/C_y)^2 < 1-(C_g/R_v)^2$ and $|R_v|>|C_g|$. 
  The latter condition can also be obtained analytically by considering the linear response of the valley Hall conductivity. Expanding Eq. \eqref{WSMham} around  $\vec{k}_0 = (0,\eta \frac{\pi}{2})$, $\eta=\pm 1$, we obtain
 \begin{equation}
L^\eta_{\mathrm{DP}} (\vec{q}) = 2 \eta  C_y q_y \sigma_x+ C_1 q_x \sigma_y + (C_g + \eta R_v) \sigma_z,
\label{eq6}
\end{equation}
where $\vec{q} \equiv \vec{k}-\vec{k}_0$. One may thus identify the two $\vec{k}_0$ points as the Dirac points (DPs) of massive Dirac fermion Hamiltonians Eq. \eqref{eq6} associated with the two valleys, so that $\eta$ takes the meaning of a valley index where  $\eta= +1 (-1)$ denotes the $K$ and $K'$ valley index, respectively. The low-admittance equation reads
\begin{equation}
\epsilon^\eta_\pm=\pm \sqrt{(2 C_y q_y)^2+(C_1 q_x )^2+(C_g + \eta R_v)^2  },
\label{eq7}
\end{equation}
where $\pm$ denotes the particle- and hole-like bands respectively. The analogous DC Hall conductivity $\sigma_{xy}$ at each valley can be calculated through the standard Kubo formula \cite{tahir2013quantum} as 
\begin{equation}
	\sigma^\eta_{xy} = \int \frac{\mathrm{d}q^2}{\pi^2} \frac{1}{(\epsilon^\eta_+-\epsilon^\eta_-)^2} \mathrm{Im}(M^{\eta+}_x(\vec{q})M^{\eta-}_y(\vec{q}))
\end{equation}
where $M^{\eta\pm}_x (\vec{q})=\langle{+},\vec{q},\eta| (\partial_{q_x}L^\eta_{\mathrm{DP}})|-,k,\eta\rangle$ and $M_y^{\eta\mp} (\vec{q})=\langle{+},\vec{q},\eta|(\partial_{q_y}L^\eta_{\mathrm{DP}}) |-,\vec{q},\eta\rangle$.  Here, $|\pm,\vec{q},\eta\rangle$ is the right eigenvector of Eq. \ref{eq6}, and and $\langle{\pm},\vec{q},\eta|$ its Hermitian conjugate. The valley-dependent DC Hall conductivity can be evaluated as 
 \begin{equation}
\sigma_{xy}^\eta=-\frac{1}{8\pi^2 } \mathrm{Sgn}(C_g + \eta R_v).
\label{eq9}
\end{equation}
From Eq. \ref{eq9}, one can see that the Hall conductivity contributions for the two valleys are unequal. We define the total quantum  valley Hall conductivity ($\sigma_{xy}^\textrm{valley} \equiv \sigma_{xy}^{\eta=1}-\sigma_{xy}^{\eta=-1}$). We obtain zero and finite $\sigma_{xy}^\textrm{valley}$ for $|C_g/R_v|>1$ and $|C_g/R_v|<1$ respectively. When $|R_v| < |C_g|$, the two valleys have the same Hall conductivities. The quantum valley Hall conductivity is hence zero, and there are no topologically non-trivial edge states which cross the bulk bandgap. We find, however, that topologically \textit{trivial} edge states still exist when boundaries are introduced along certain directions but these do not cross the bandgap.

 \begin{figure*}[htp!]
  \centering
    \includegraphics[width=0.9\textwidth]{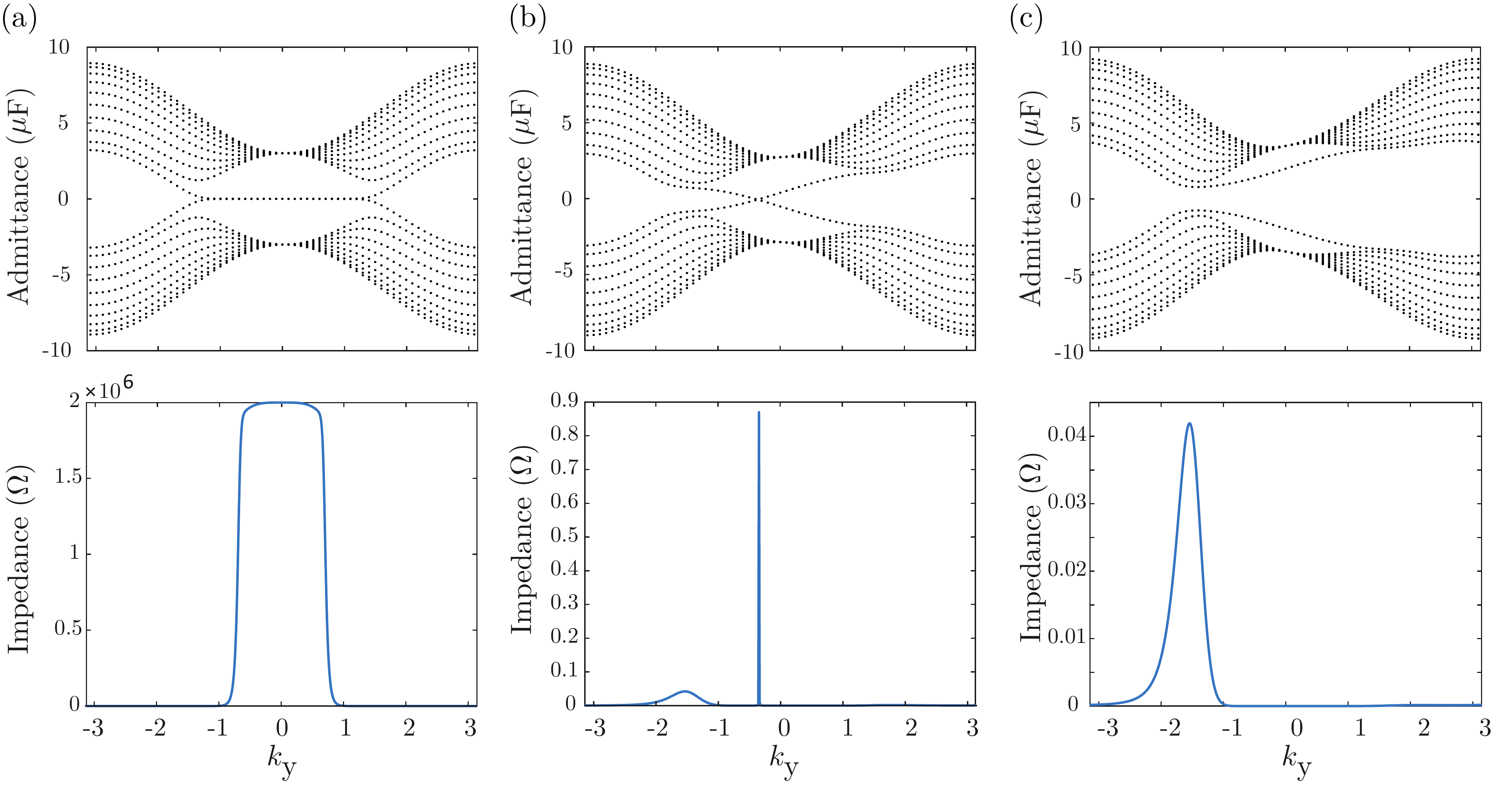}
  \caption{Simulated admittance and impedance profiles of the valley TE circuit. The simulation is performed via the electric circuit simulation software LTspice. The top rows show the admittance dispersion relations of $x$-confined nanoribbons with 10 unit cells along the $x$ direction at the resonant frequency $f=25.250\ \mathrm{Hz}$. For a realistic simulation, the components are selected from the LTspice component database i.e.,  $C_1=3\ \mu \mathrm{F}$ ($2 \times$ Murata GRM033R60G155ME14D), and $C_y=1.5\ \mu \mathrm{F}$ (Murata GRM033R60G155ME14D) for (a) edge states with $C_g=0\ \mu \mathrm{F}$ and $R_v= 0$, (b) the non-trivial topological phase $C_g=0.56\ \mu \mathrm{F}$ (KEMET C1206C564K3RACTU) and $R_v=2$, and (c) the trivial valley phase with $C_g=2.2\ \mu \mathrm{F}$ (KEMET C0603C225K9PAC) and $R_v=0.5$. For the common grounding inductors, $L=10\  \mu \mathrm{H}$ (W\"{u}rth Elektronik 744042100) is used. To realize the negative resistive element $R_v$, we performed the simulation with the high precision operational amplifier LT1056. The bottom rows show the respective spectra of the impedances measured between the two edges of the nanoribbons.}
  \label{fig2}
\end{figure*}

Fig. \ref{gphasesFig}c shows the dispersion relations at representative points on the $R_v-C_1$ plane for nanoribbon geometries of the TE circuit in which the circuits have infinite length along the $x$ direction and 10 unit cells in the $y$ direction (left), and in which the circuits have infinite length along the $y$ direction and 10 unit cells in the $x$ direction (right). (The admittance dispersion is the TE analogue of the energy dispersion in quantum mechanical systems.) In general, the dispersion relations for the finite $x$ and finite $y$ nanoribbons appear markedly different from one another although their topological character remains unchanged when the finite direction is exchanged. Points (i) and (iii) on Fig. \ref{gphasesFig}b exemplify the $|R_v|<|C_g|$ scenario, with $|C_1| > |C_y|$ at point i and $|C_1| < |C_y|$ at point (iii). At both of these points, no edge states exist when the nanoribbon confinement direction is along the $y$ direction. However, edge states emerge when the nanoribbon confinement is along the $x$ direction. The edge states are more prominent at $|k_y| < \pi/2$ in the case of $|C_1| > |C_y|$ corresponding to  point (i) whereas the edge states confined to a narrow range of $k_y$  in the vicinity of $|k_y|=\pm \pi/2$ in the case of $|C_1|< |C_y|$ corresponding to  point (iii). Point (iv) corresponds to the scenario of $|R_v| > |C_g|$, $(C_y/C_1)^2 > 1-(C_g/R_v)^2$ scenario. In contrast to points (i) and (iii) discussed previously, edge states appear on point (iv) only when the nanoribbon confinement direction is in the $x$ direction, but not in the $y$ direction. Note that the edge states corresponding to (i), (iii) and (iv) do not cross the bandgap, thus indicating their trivial character. Finally, point (ii) exemplifies the topologically non-trivial phase which exists when $|R_v| > |C_g|$, $(C_y/C_1)^2 < 1-(C_g/R_v)^2$. Topologically protected edge states which cross the bulk gap are present for both nanoribbon confinement directions. In the remainder of this paper, we will focus on the phases exemplified by points (i) (topologically trivial) and (ii) (topologically non-trivial) for the case of $|C_1| > |C_y|$, and without loss of generality set $C_1 = 2C_y$.

To further investigate the effect of $C_g$ and $R_v$ on the TE model when $|C_1| > |C_y|$, we simulated the circuit in LTspice with realistic device parameters and plot the admittance and impedance spectra as functions of wavevector $k_y$ for open boundary conditions along the $x$ direction (see Fig. \ref{fig2}a-c). The impedance between any two lattice sites $p$ and $q$ in the TE network model is given by
\begin{equation}
\begin{aligned}
Z_{pq}= & (L_{TE})_{pp}^{-1}-(L_{TE})_{pq}^{-1}+(L_{TE})_{qq}^{-1}-(L_{TE})_{qp}^{-1} \\
	   =&\sum_{k=1}^N \frac{{|\phi_{kp}-\phi_{kq}|}^2}{\epsilon_k},
\end{aligned}
\label{eq5}
\end{equation}
where $\phi_{ij}$ is value of the $i^{th}$ eigenvector at the $j^{th}$ lattice point and $\epsilon_i$ is the $i^{th}$ non-singular eigenenergy of the Laplacian matrix. Note that LTspice simulations took into account tolerance windows and parasitic effects. Even after including such effects/imperfections, the topological behaviour remains largely  unaffected.   

In the upper plot in Fig. \ref{fig2}a, we find  well defined edge states for zero $C_g$ and $R_v$. Moreover, the whole admittance spectra is symmetric about the zero admittance line. This is due to the chiral symmetry of the Laplacian $\mathcal{C}L_{TE}(\vec{k})\mathcal{C}^{-1}=-L_{TE}(\vec{k})$ where $\mathcal{C} = \sigma_z$ is the chiral inversion operator.  However, the boundary modes evolution would be different for finite mass term  i.e. ($C_g + \eta R_v$) in Eq. \eqref{eq6}. When $|C_g| < |R_v|$, the gapless edge modes emerge (see Fig. \ref{fig2}b), where the zero-energy edge states split into two  tilted boundary states that intersect each other. However when the resistive coupling becomes stronger than the onsite capacitor (i.e., $|C_g| > |R_v|$ ), gapped boundary modes appear, as  shown in Fig. \ref{fig2}c. Interestingly, a finite mass term breaks the symmetry of admittance dispersion about the admittance $E=0$ line (Fig. \ref{fig2}b,c).
The impedance spectra for Fig. \ref{fig2}a--c are shown in the lower plots of the corresponding panel. The nearly zero-admittance edge states in Fig. \ref{fig2}a are marked by the very large impedance for $|k_y|\lesssim \pi/2$ , which agrees with the inversely proportional relation between the eigenvalue (admittance) and the impedance in Eq. \ref{eq5}. However, the impedance falls significantly in presence of a mass term in the circuit Laplacian and only discrete impedance peaks are found when the admittance gap between the two bands reaches close to zero (see Fig. \ref{fig2}b, c). The gapless and gapped edge states are indicated by the comparatively large and small impedance peaks respectively (compare the sharp peak in Fig. \ref{fig2}b with the broader lower peak of Fig. \ref{fig2}c).   The close correspondence between the admittance spectra and impedance readouts obtained from  the LTspice simulation  demonstrates the experimental realization of valley-dependent features and their electrical characterization under realistic conditions.  

\section{Valley kink states}
\begin{figure*}[ht]
  \centering
\includegraphics[width=0.85\textwidth]{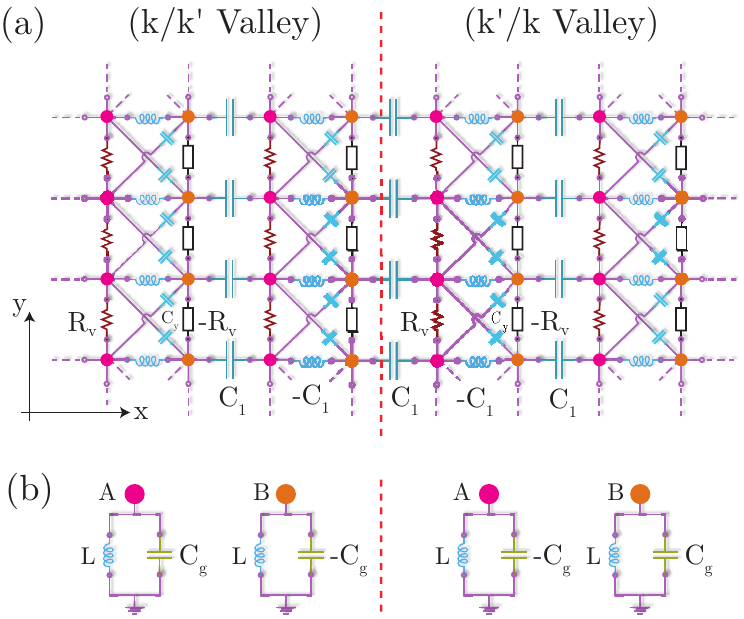}
  \caption{ Schematic of a TE circuit heterojunction that exhibits kink states. (a) Valley kink circuit. (b) Schematic of the wiring between each node and the ground in the two regions. Note that the onsite potential parameter (i.e., $C_g$) changes sign between the two regions. This guarantees that the Hall conductivity of each of the valleys switches sign across the heterojunction interface. All the other parameters are the same in both regions.  }
  \label{fig6}
\end{figure*} 
In the previous section, we studied the relation between the valley-dependent topological phases and the emergence of edge states. In this section, we study the evolution of the boundary states evolution in a heterojunction between two TE segments in which the Hall conductivities of each valley have opposite signs in the two segments. More explicitly, we realize a topolectrical valley kink state at the domain wall-like interface between two TE circuit arrays with opposite signs of the quantum valley Hall conductivities as shown in Fig. \ref{fig6}. Such heterojunctions can be realized by simply reversing the sign of the onsite capacitance $C_g$ between the two array segments for a fixed $|R_v|<|C_g|$. The TE lattice in
the left and right sides of the Interface have the same model parameters except for the onsite potential $C_g$, which is positive and negative on the left and right sides respectively. Such a sign change in one of  the mass parameter terms(i.e., $C_g$) across the heterojuction will induce valley-dependent kink states localized at the interface. Here we focus on heterojunctions where the interface is parallel to the $y$ direction. 
\begin{figure*}[htp!]
  \centering \includegraphics[width=0.90\textwidth]{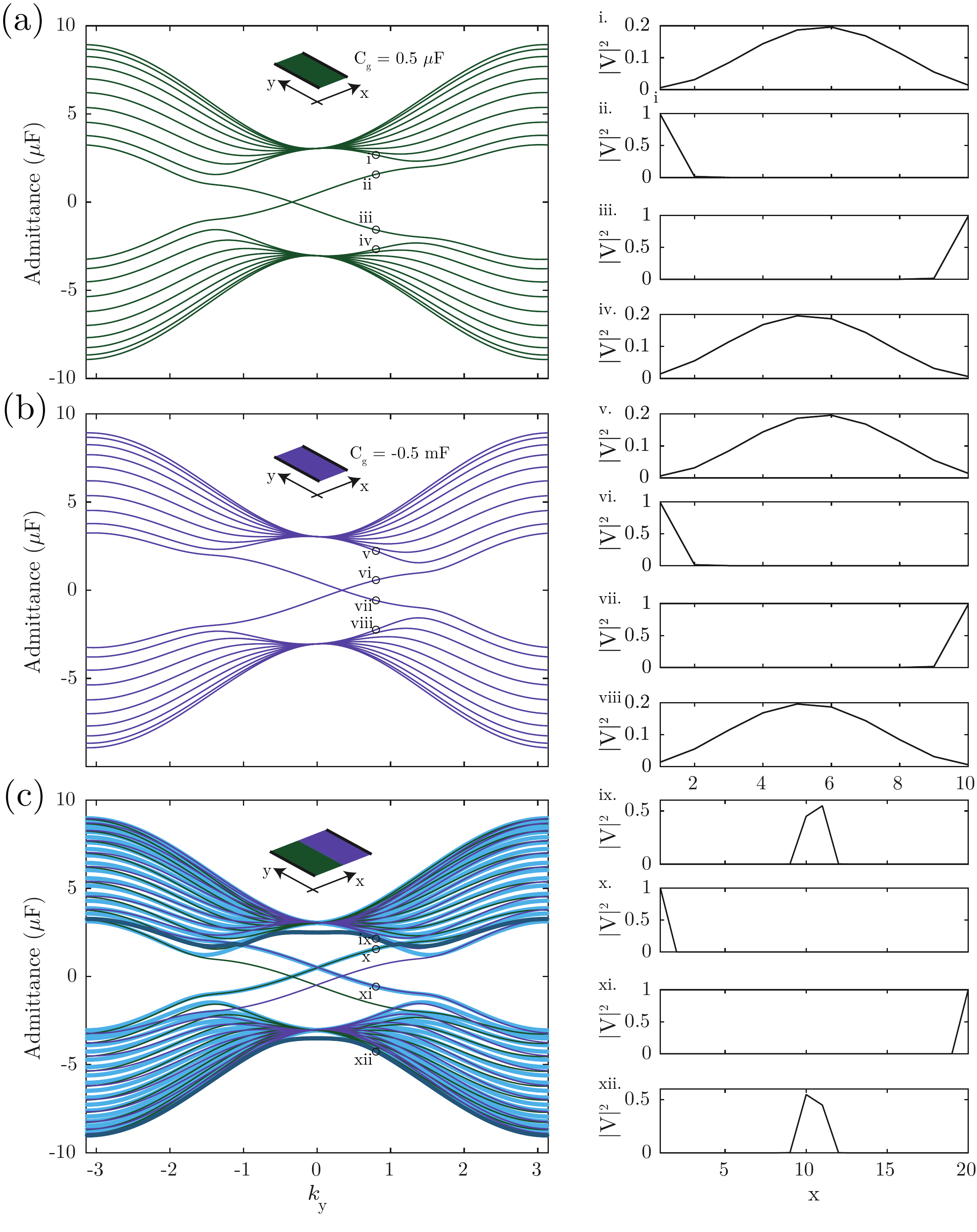}
  \caption{The plots on the left show the admittance dispersions for TE circuits with $C_1 = 3\ \mu \mathrm{F}$, $R_v=0.75\ \mu \mathrm{F}$, $C_y=1.5\ \mu \mathrm{F}$ for (a) a 10 unit cells-wide $C_g=0.5\ \mu \mathrm{F}$ nanoribbon and (b) a 10 unit cells-wide $C_g=-0.5\ \mu \mathrm{F}$ nanoribbon. The thicker lines in panel (c) show the dispersion relation of the heterojunction consisting of the nanoribbon in (a) on the left and the nanoribbon in (b) on the right connected together along the $y$ direction with the dispersion relations of the isolated left and right nanoribbons superimposed as thinner lines. The thicker dark lines denote the valley kink states. The plots on the right indicate the voltage amplitudes of the bands labelled i to xii in the dispersion relations at $k_y=0.8$.  }
  \label{gkink1}
\end{figure*} 

Fig. \ref{gkink1} shows the admittance dispersion of the left (Fig. \ref{gkink1}a) and right (Fig. \ref{gkink1}b) halves of the heterojunction (Fig. \ref{gkink1}c) in isolation from each other, and that of the entire heterojunction when both halves of the heterojunction are in the topologically non-trivial phase. In the particular heterojunction under consideration,, the TE circuit array to the left of the heterojunction has a positive $C_g$ while the right half has a negative $C_g$ of the same magnitude. One way of ascertaining the topological character of a particular state is to plot its spatial distribution. Hence, the square of the voltage amplitudes summed over the A and B nodes in each unit cell are plotted for some of the bands as a function of the $x$ coordinate  across the transverse width of the TE. The square of the voltage amplitude constitutes the TE analogues of the quantum mechanical probabilty densities $\psi_x^\dagger\psi_x$ for states described by the wavefunction spinor $\psi_x$, and shall be loosely referred to as `probability densities' for brevity henceforth. The probability densities in panels (a) and (b) show that the bands crossing the admittance bandgap (ii, iii, vi, and vii) consist of edge states exhibiting edge localization. Furthermore, when the  edge states have positive (negative) $k_y$ slopes, i.e., ii and vi (iii and vii), the corresponding states   are localized on the left (right) edges of the nanoribbon. The other bands labelled as i, iv, v, and viii consist of bulk states where the highest probability densities occur near the center of the nanoribbon and away from the edges. Comparing Fig.  \ref{gkink1}a and  \ref{gkink1}b, we find that the $k_y$ dispersion  of  nanoribbons with opposite signs of $C_g$ are reflections of each other about the $k_y$ axis. This reflection symmetry can be understood from the form of the Laplacian Eq. \eqref{WSMham}: the Laplacian is invariant upon the simultaneous replacement of $C_g \rightarrow -C_g$ and a reflection about the $x$ axis, which brings $k_x \rightarrow -k_x$ and $\sigma_{y,z} \rightarrow -\sigma_{y,z}$. 

When the two halves of the heterojunction are connected together in the heterojunction, the resultant dispersion relation of the bulk bands is roughly given by the superposition of the bulk band dispersions in the isolated halves of the heterojunction, as shown in Fig.~\ref{gkink1}c. There is a slight increase in the energy separation between corresponding pairs of bulk bands from each of the isolated halves in the heterojunction due to the band anticrossing. Note that the edge states localized away from the heterojunction interface in either half of the TE heterojunction circuit are not significantly perturbed when the heterojunction is formed. For instance, the dispersion and spatial distribution of  band xi (x) is virtually identical to that of  band vii (ii) which is localized at the right (left) edge of the isolated half circuits. In contrast, the edge states of the isolated halves localized at the edges adjacent to the heterojunction interface (iii and viii) have disappeared in the heterojunction. They are replaced by bands ix and xii that emerge in the TE heterojunction circuit which do not have corresponding counterparts in the isolated halves of the heterojunctions circuit. These new bands correspond to the valley kink states localized at the heterojunction interface as depicted in the spatial variation of $|V|^2$ in Fig. \ref{gkink1}(c).  
\begin{figure*}[htp!]
  \centering
    \includegraphics[width=0.9\textwidth]{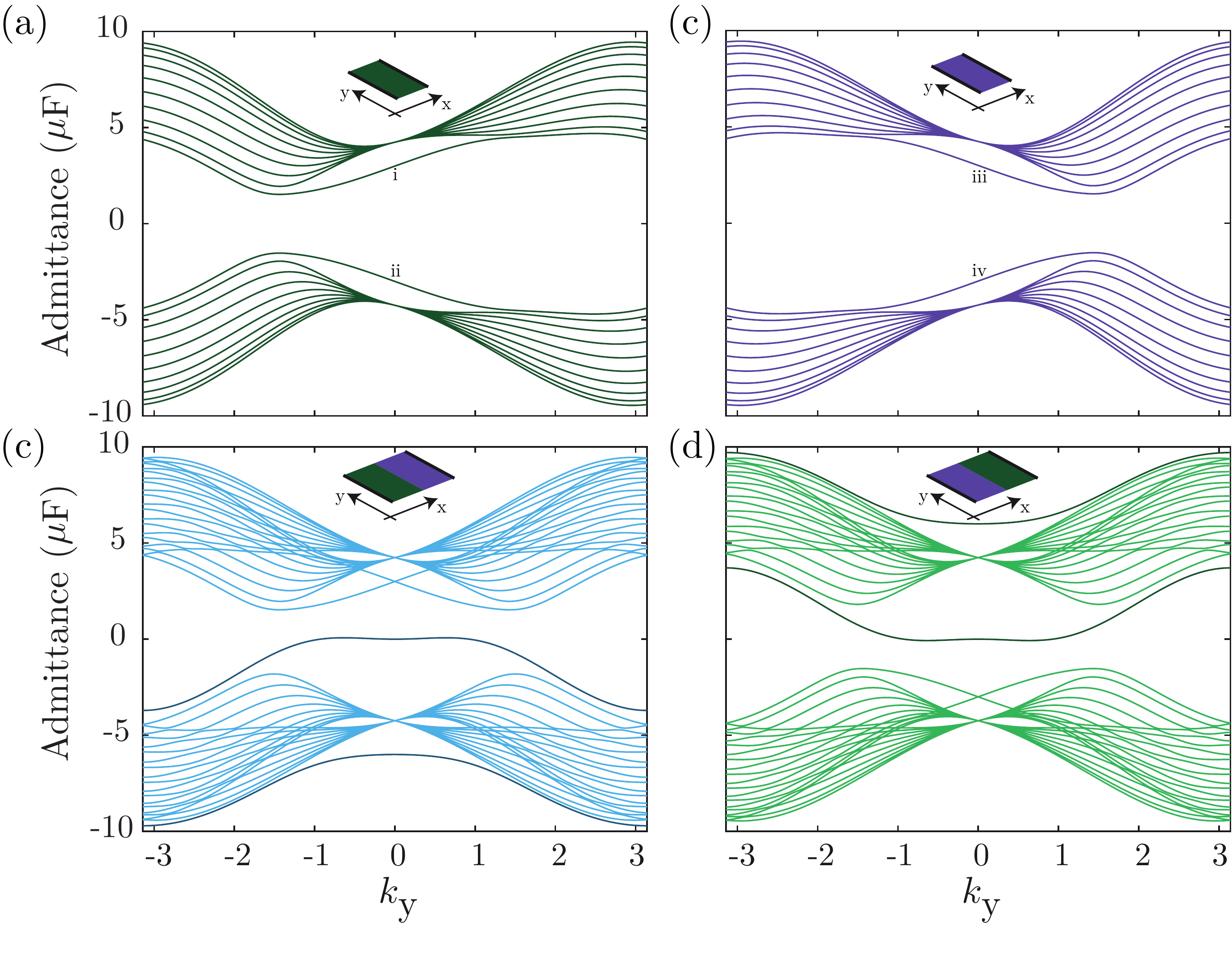}
  \caption{(a) and (b) show the band dispersions of 10 unit cells-wide TE nanoribbons with $C_1 = 3\  \mu \mathrm{F}$, $R_v=0.75\ \mu \mathrm{F}$, $C_y=1.5\ \mu \mathrm{F}$ and (a)  $C_g=3\ \mu \mathrm{F}$ and (b) $C_g=-3\ \mu \mathrm{F}$. (c) shows the dispersion relation of the heterojunction formed when the $C_g=3\ \mu \mathrm{F}$ nanoribbon is on the left and the $C_g=-3\ \mu \mathrm{F}$ nanoribbon on the right, and (d) that of the heterojunction when the $C_g=-3\ \mu \mathrm{F}$ nanoribbon is on the left and the $C_g=3\ \mu \mathrm{F}$ nanoribbon on the right. The valley kink states are indicated as the darker lines in panels (c) and (d).  }
  \label{ggappedKink}
\end{figure*} 

Surprisingly, valley kink states can also be formed in heterojunctions formed from joining together two topologically trivial nanoribbons  segments, i.e.,  each with $|C_1| > |C_y|$, as shown in Fig. \ref{ggappedKink}. Panels (a) and (b) show the dispersion relations for the isolated nanoribbons with positive and negative values of $C_g$, respectively. The particle-like (hole-like) edge states with positive (negative) $k_y$ slopes, i.e. i and iii (ii and iv) are localized at the left (right) edges of the isolated nanoribbons. The dispersion relations and the energies of the kink states in the heterojunction depend on the signs of $C_g$ in the two halves of the heterojunction. When the positive (negative) $C_g$ nanoribbon is on the left (right) half of the heterojunction, the particle-like edge bands of both the left and right isolated halves (i.e. bands (i) and (iii)) which are localized away from the heterojunction interface are still be preserved in the heterojunction circuit, while the resultant valley kink states are hole-like (i.e. bands (ii) and (iv)), as shown in panel Fig. \ref{ggappedKink}(c). Conversely, when the negative (positive) $C_g$ nanoribbon is on the left (right) half of the heterojunction, the hole-like edge states of the isolated halves are preserved in the heterojunction circuit while the resultant valley kink states would be particle-like. 
\begin{figure*}[htp!]
  \centering
    \includegraphics[width=0.9\textwidth]{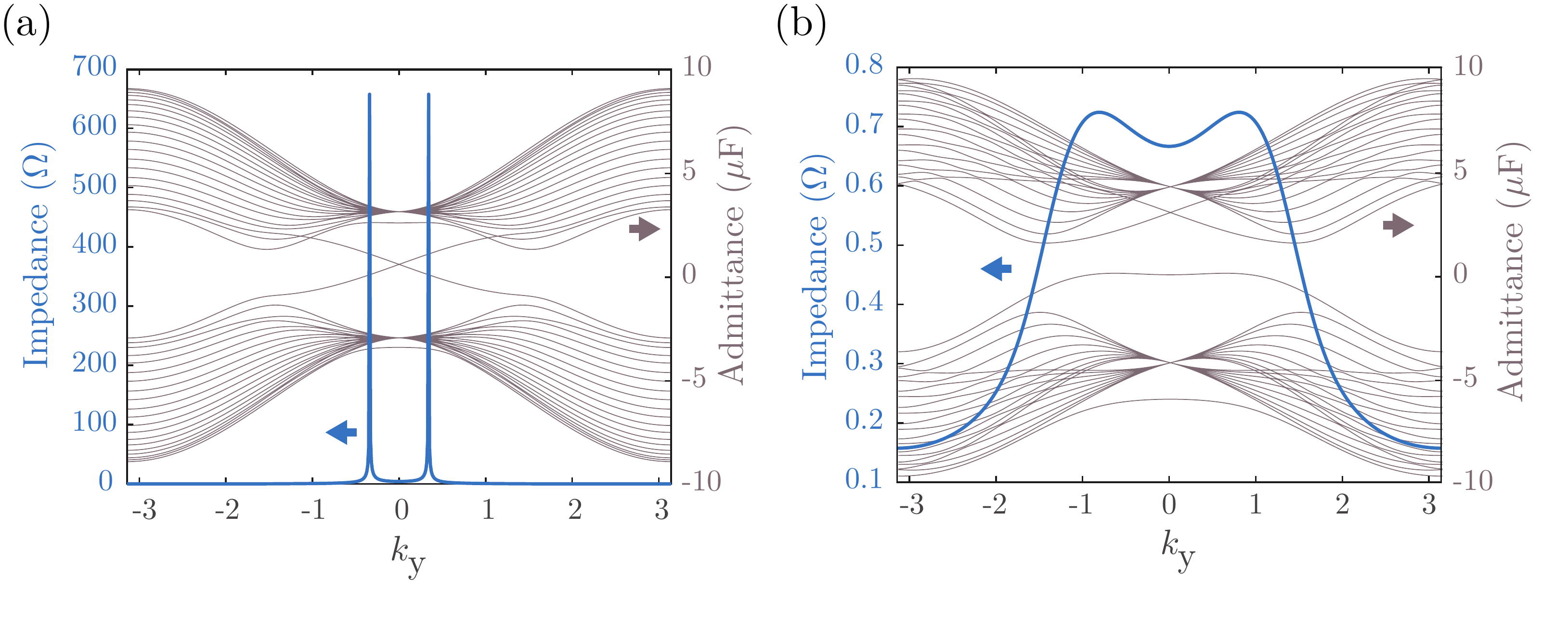}
  \caption{(a) and (b) show the band dispersions (thin lines) and the terminal impedance between the leftmost and rightmost nodes of the gapless heterojunction in Fig. \ref{gkink1}c and the gapped heterojunction in \ref{ggappedKink}c respectively.  }
  \label{gkinkImpPlots}
\end{figure*} 

Finally, the heterojunction circuit with the presence of valley kink states can be characterized by their unique impedance signatures, as shown in Fig.~\ref{gkinkImpPlots}. Unlike the isolated nanoribbons which  exhibit only a single impedance peak as a function of $k_y$  across the Brillouin zone, the impedance dispersion of the heterojunction circuit has a $k_y$-reflection symmetry which leads to a pair of impedance peaks. The impedance peaks in the impedance dispersion of the  gapped heterojunction circuit are located near the points the hole- and particle-like bands have their minimum energy separation  (see Fig. \ref{gkinkImpPlots}b), similar to the dispersion of isolated nanoribbon in Fig.~\ref{fig2}c. In contrast, the impedance peaks of the gapless heterojunction circuit are displaced from $k_y=0$ where the hole- and particle-like bands meet. Furthermore, the peak impedance of the gapless heterojunction is much higher than its  gapped heterojunction counterpart. 
\section{Conclusion}
In summary, we proposed a highly tunable TE platform that exhibits topological valley Hall states and valley kink modes. The circuit can be switched between different topological valley Hall states simply by varying the sign of the onsite capacitance. Moreover, gapped and gapless boundary states emerge in the admittance spectra with proper tuning of the mass parameters.  We also realize valley kink states by connecting two TE circuits with opposite signs of Hall conductivites together. We analytically derived the boundary conditions between trivial and no-trivial states, characterized by their Chern number. The gapped and gapless states can be switched from one to the other by tuning the grounding capacitance. The topological boundary modes or valley kink states are localized in the interface of the kink circuit.  There are significant differences in the impedance dispersion for the different boundary modes, leading to  measurable and distinguishable circuit responses for different topological valley and kink states in the uniform and heterojunction TE circuits, respectively. In summary, our work based on the TE circuit model provides an accessible testbed to realize various topological valley and kink phases  and allows the efficient modulation and switching between different topological states for valleytronic applications.


%

\end{document}